\documentclass[conference]{IEEEtran}
\IEEEoverridecommandlockouts
\usepackage{cite}
\usepackage{amsmath,amssymb,amsfonts}
\usepackage{algorithmic}
\usepackage{graphicx}
\usepackage{textcomp}
\usepackage{xcolor}
\usepackage{tikz}
\usepackage{url,hyperref}

\def\BibTeX{{\rm B\kern-.05em{\sc i\kern-.025em b}\kern-.08em
    T\kern-.1667em\lower.7ex\hbox{E}\kern-.125emX}}
\definecolor{rosso}{RGB}{220,57,18}
\definecolor{giallo}{RGB}{255,153,0}
\definecolor{blu}{RGB}{102,140,217}
\definecolor{verde}{RGB}{16,150,24}
\definecolor{viola}{RGB}{153,0,153}

\makeatletter

\tikzstyle{chart}=[
    legend label/.style={font={\scriptsize},anchor=west,align=left},
    legend box/.style={rectangle, draw, minimum size=5pt},
    axis/.style={black,semithick,->},
    axis label/.style={anchor=east,font={\tiny}},
]

\tikzstyle{bar chart}=[
    chart,
    bar width/.code={
        \pgfmathparse{##1/2}
        \global\let\bar@w\pgfmathresult
    },
    bar/.style={very thick, draw=white},
    bar label/.style={font={\bf\small},anchor=north},
    bar value/.style={font={\footnotesize}},
    bar width=.75,
]

\tikzstyle{pie chart}=[
    chart,
    slice/.style={line cap=round, line join=round, very thick,draw=white},
    pie title/.style={font={\bf}},
    slice type/.style 2 args={
        ##1/.style={fill=##2},
        values of ##1/.style={}
    }
]

\pgfdeclarelayer{background}
\pgfdeclarelayer{foreground}
\pgfsetlayers{background,main,foreground}

\newcommand{\pie}[3][]{
    \begin{scope}[#1]
    \pgfmathsetmacro{\curA}{90}
    \pgfmathsetmacro{\r}{1}
    \def\c{(0,0)}
    \node[pie title] at (90:1.3) {#2};
    \foreach \v/\s in{#3}{
        \pgfmathsetmacro{\deltaA}{\v/100*360}
        \pgfmathsetmacro{\nextA}{\curA + \deltaA}
        \pgfmathsetmacro{\midA}{(\curA+\nextA)/2}

        \path[slice,\s] \c
            -- +(\curA:\r)
            arc (\curA:\nextA:\r)
            -- cycle;
        \pgfmathsetmacro{\d}{max((\deltaA * -(.5/50) + 1) , .5)}

        \begin{pgfonlayer}{foreground}
        \path \c -- node[pos=\d,pie values,values of \s]{$\v\%$} +(\midA:\r);
        \end{pgfonlayer}

        \global\let\curA\nextA
    }
    \end{scope}
}

\newcommand{\legend}[2][]{
    \begin{scope}[#1]
    \path
        \foreach \n/\s in {#2}
            {
                  ++(0,-10pt) node[\s,legend box] {} +(5pt,0) node[legend label] {\n}
            }
    ;
    \end{scope}
}

\begin{document}

\title{Lest We Forget:  A Dataset of Coronavirus-Related  News Headlines in Swiss Media
}

\author{\IEEEauthorblockN{Alireza Ghasemi}
\IEEEauthorblockA{\textit{ELCA Informatik AG} \\
Z\"urich, Switzerland \\
alireza.ghasemi@elca.ch}
\and
\IEEEauthorblockN{Amina Chebira}
\IEEEauthorblockA{\textit{ELCA Informatique SA} \\
Lausanne, Switzerland \\
amina.chebira@elca.ch}
}

\maketitle

\begin{abstract}
We release our COVID-19 news dataset, containing more than 10,000 links to news articles related to the Coronavirus pandemic published in the Swiss media since early January 2020. This collection can prove beneficial in mining and analysis of the reaction of the Swiss media to the COVID-19 pandemic and extracting insightful information for further research.

We hope this dataset helps researchers and the public deliver results that will help analyse the pandemic and potentially lead to a better understanding of the events.
\end{abstract}


\section{Introduction}
The COVID-19 pandemic \emph{started} in Switzerland on February 25\textsuperscript{th} 2020, when the first infection was officially reported in the Italian-speaking canton of Ticino~\cite{FirstCase:EN,FirstCase:DE}. Soon the pandemic spread around the country on all cantons, and Switzerland became one of the most infected countries on a per-capita basis~\cite{Vaud:DE,Spread:EN}. 

The Swiss government started putting in place various measures to control and suppress the pandemic. Gatherings were limited and later totally banned, following by closure of all except essential business, and finally closing land borders with neighbouring countries~\cite{Measures:DE}. 

These measures helped control the spread of the virus and significantly decreased the number of active and daily new cases in Switzerland. With the success confirmed, government started gradually lifting the established restrictions from late April~\cite{Lifting}. Finally, the June 15\textsuperscript{th} re-opening land borders with the neighbouring countries marked the \emph{"end"} of the pandemic in Switzerland, at least for the first wave~\cite{BorderOpening}. 

Since the first recorded case of the COVID-19 virus in Switzerland and far before as it was gaining attention around the world, The Swiss media started covering the topic from various aspects, including the everyday news about the state of the country, the immediate effects, and longer-term consequences of the pandemic. Given the multi-lingual and multi-cultural nature of Switzerland, interesting analyses can be accomplished to see how the media coverage of the pandemic has been managed and what topics in respect to the pandemic have been important to the Swiss media, and hopefully, by proxy to the Swiss public. 

In order to help the research community and the public be able to analyse and seek answers to the above questions, we at ELCA decided to release our COVID-19 news dataset, containing more than 10,000 links to news articles related to the Coronavirus pandemic published in the Swiss media since early January 2020. 

We hope this dataset helps researchers make insightful analyses on the reaction of the Swiss public to the pandemic and deliver results that help shape a better response in the prospective future cases.

\section{The Data}
We tried to cover the most popular Swiss newspapers and news websites. Therefore, we chose a total of 10 news sources in German, five in French, three in Italian, and also two English-speaking Swiss news websites in order to make the data more accessible to to researchers outside Switzerland. Table \ref{tab11} depicts the list of venues and some information about them.

\begin{table}[htbp]
\caption{Selected News Sources}
\begin{center}
\begin{tabular}{|c|c|c|}
\hline
\textbf{News Source} & \textbf{Language}& \textbf{Number of Articles} \\
\hline 
Blick     & German &                2080 \\ \hline
Neue Zürcher Zeitung & German &       1961 \\ \hline
Tages Anzeiger  & German &            1109 \\ \hline
Aargauer Zeitung & German &           1099 \\ \hline
Basler Zeitung & German &             859  \\ \hline
SRF & German &           718   \\ \hline
20 Minuten & German &            553 \\\hline
Berner Zeitung  & German &            281 \\\hline
20 Minutes & French &            225 \\\hline
Tribune de Genève & French &          223 \\\hline
Corriere del Ticino & Italian &        199 \\\hline
24 heures  & French &                 189 \\\hline
World Radio Switzerland & English &     176 \\\hline
Il portale del Ticino  & Italian &     156 \\\hline
RTS & French &             188 \\\hline
SWI swissinfo.ch & English &      112 \\\hline
Le Temps & French &     89 \\\hline
SWI swissinfo.ch  & French &  71 \\\hline
SWI swissinfo.ch  & German &       55 \\\hline
SWI swissinfo.ch  & Italian &       52 \\\hline
\end{tabular}
\label{tab11}
\end{center}
\end{table}

\subsection{Selection of Relevant Articles}
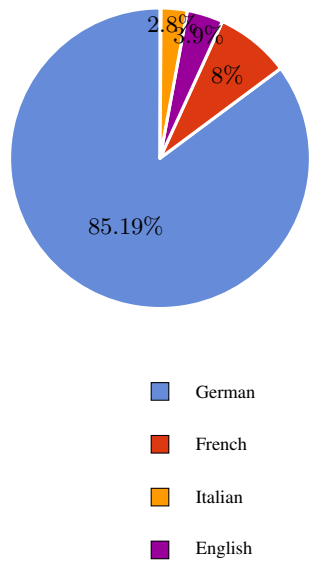
\begin{figure}
    \centering
    \begin{tikzpicture}
    [
        pie chart,
        slice type={comet}{blu},
        slice type={legno}{rosso},
        slice type={coltello}{giallo},
        slice type={sedia}{viola},
        slice type={caffe}{verde},
        pie values/.style={font={\small}},
        scale=2
    ]
    
        \pie{}{85.19/comet,8/legno,3.9/sedia,2.8/coltello}

        \legend[shift={(0cm,-1.2cm)}]{{German}/comet, {French}/legno,  {Italian}/coltello,  {English}/sedia}

\end{tikzpicture}
    \caption{Language Distribution of Articles in the Dataset}
    \label{fig:my_label}
\end{figure}

We backdated our data collection to late 2019, and started scanning front pages of the selected news sources in consecutive days, extracting headlines of the articles. Initially, articles with any of the following \emph{keywords} in the title were deemed \emph{"Coronavirus-related"}:
\begin{itemize}
\item Coronavirus,
\item Covid,
\item Lockdown,
\item Pandem* (To account for different spellings of the concept in different languages).
\end{itemize}
This inevitably leads to false negatives. In order to reduce such false negatives, we read at a later stage the synopsis of the article and searched for the keywords also in the body, yielding more positive results. The distribution of the languages in the dataset is depicted in Figure \ref{fig:my_label}.

The first article we could find in the Swiss media has been published on January 8\textsuperscript{th} in the French-speaking news portal \emph{20 Minutes}, titled \emph{"A new Coronavirus appears in China"}~\cite{FirstArticle}. We have made a web application to simplify exploring and browsing the data, and reading the collected news articles. The web application is available at \url{https://covidnewsdataset.herokuapp.com/}.
\section*{Summary}

We explained in this article our Swiss COVID-19 dataset and how it has been collected. We publish the dataset hereby for public use, along with an online visualisation application to help explore and look at the news articles of Swiss media during the pandemic in Switzerland. We hope this dataset proves useful in analysis of the pandemic era and the public response to it in Switzerland. 

\def\UrlBreaks{\do\/\do-}
\Urlmuskip=0mu plus 1mu\relax
\bibliographystyle{./bibliography/IEEEtran}
\bibliography{./bibliography/IEEEexample}

\end{document}